\documentstyle[aps,floats,psfig]{revtex}

\begin{document}
\psfigurepath{.:plot:figure}
\twocolumn[\hsize\textwidth\columnwidth\hsize\csname @twocolumnfalse\endcsname
\bibliographystyle{unsrt}
\preprint{draft:comm6.tex, \today}
\title{\bf Commensurate dynamic magnetic correlations in 
La$_2$Cu$_{0.9}$Li$_{0.1}$O$_4$}
\author{Wei Bao,$^{1}$ R. J. McQueeney,$^{1}$ R. Heffner,$^{1}$ 
J. L. Sarrao,$^{1}$
P. Dai,$^{2}$ J. L. Zarestky$^{3}$}
\address{$^{1}$Los Alamos National Laboratory, Los Alamos, NM 87545\\
$^{2}$Oak Ridge National Laboratory, Oak Ridge, TN 37831\\
$^{3}$Ames Laboratory, Ames, IA 50011
}
\date{\today}
\maketitle
\begin{abstract}
When sufficient numbers of holes are introduced into the two-dimensional
CuO$_2$ square lattice, dynamic magnetic correlations become 
{\em in}commensurate
with underlying lattice in all previously investigated
La$_{2-x}$$A_x$Cu$_{1-z}$$B_z$O$_{4+y}$ ($A=$ Sr or Nd, $B=$ Zn)
including high $T_C$ superconductors and insulators, 
and in bilayered
superconducting YBa$_2$Cu$_3$O$_{6.6}$ and Bi$_2$Sr$_2$CaCu$_2$O$_8$. 
Magnetic correlations also become {\em in}commensurate in
structurally related La$_{2}$NiO$_{4}$ when doped with Sr or O.
We report an
exception to this so-far well established experimental ``rule'' in
La$_2$Cu$_{1-z}$Li$_{z}$O$_4$ in which magnetic correlations remain
commensurate.
\end{abstract}
\vskip2pc]

\narrowtext

High transition temperature ($T_C$) superconductivity is realized 
when charge carriers are introduced to the CuO$_2$ planes of 
the insulating parent compound, for example,
La$_2$CuO$_4$ or YBa$_2$Cu$_3$O$_{6}$. 
These parent compounds are now well-established as two-dimensional (2D)
spin $S=\slantfrac{1}{2}$ Heisenberg antiferromagnets with a dominant in-plane 
exchange interaction\cite{la3dv,la2dv,la2smhao,la2cl}.
The evolution of the magnetic correlations with charge
carrier doping is a central issue in high $T_C$ superconductivity
research. It has so far been most extensively
investigated in the La$_2$CuO$_4$ system, due to the availability of large
single crystals. These studies show that, 
when the long-range antiferromagnetic order is suppressed and the doped
hole concentration exceeds about 5\%, 
incommensurate dynamic magnetic correlations develop
at a quartet of wave vectors 
${\bf Q}=(\slantfrac{1}{2}\pm \delta, \slantfrac{1}{2},0)$ and 
$(\slantfrac{1}{2}, \slantfrac{1}{2}\pm \delta,0)$\cite{la2ch,la2hy,la2tt}. 
Even more remarkable is that the incommensurability $\delta$ is
a universally increasing function of the hole concentration $n$\cite{L214_ym},
whether the doped material is superconducting or insulating\cite{insu} and
whether dopant resides on the La\cite{L214_Sr,L214_Nd},
the Cu\cite{L214_Zn}, or the oxygen sites\cite{L214_O}. 
In the isostructural insulating nickelates, the
magnetic correlations are also found
to be incommensurate when a sufficient number of holes are introduced by 
doping either the La or the O sites\cite{Ni214_Sr}. 
Recently, incommensurate magnetic
correlations were also discovered in bilayered superconducting 
YBa$_2$Cu$_3$O$_{6.6}$\cite{ybco_dai}. The incommensurability 
$\delta$ vs.\ $n$ falls on the same curve as the La$_2$CuO$_4$ system,
adding new excitement to this field. There is also experimental evidence
indicating incommensurate magnetic correlations in superconducting
Bi$_2$Sr$_2$CaCu$_2$O$_8$\cite{bssco}.
Empirically, therefore, there is an overwhelming consensus that doping the 
2D antiferromagnet on the CuO$_2$ square lattice eventually makes 
magnetic correlations incommensurate.
There are currently several competing theoretical
explanations for the origin of this incommensurability, 
ranging from stripe models\cite{stripe_jt,stripe0,stripe_ve0,stripe1}
where charge carriers undergo phase separation, to nesting Fermi surface 
models\cite{nfs_si} and doped quantum antiferromagnet 
models\cite{sq2dv}
where the carriers remain uniform in the sample.   

La$_2$Cu$_{1-z}$Li$_z$O$_4$ remains an insulator for $0\le z\le0.5$.
It has identical in-plane lattice parameters as Sr-doped La$_2$CuO$_4$ at 
the same hole concentration\cite{Li214}. The long-range antiferromagnetic
order is similarly suppressed by Sr or Li doping\cite{Li214,Li214muNR}.
The spin dynamics of La$_2$Cu$_{1-z}$Li$_z$O$_4$, as revealed by {\em local}
dynamical probes such as nuclear quadrapole
resonance (NQR), show an astonishing parallel with
La$_{2-x}$Sr$_x$CuO$_4$\cite{Li214NQR}. 
This suggests a similar temperature dependence in the low energy 
spin fluctuations.
In contrast to the enormous empirical conformity among 
charge-doped laminar cuprates, 
however, as will be presented
below, La$_2$Cu$_{1-z}$Li$_z$O$_4$ is 
exceptional in that the dynamic magnetic correlations remain commensurate
with the square lattice. 
This experimental result thus
provides a new facet of the rich physics 
relating antiferromagnetism and charge correlations
in charge-doped cuprates. 

\begin{figure}[bt]
\centerline{
\psfig{file=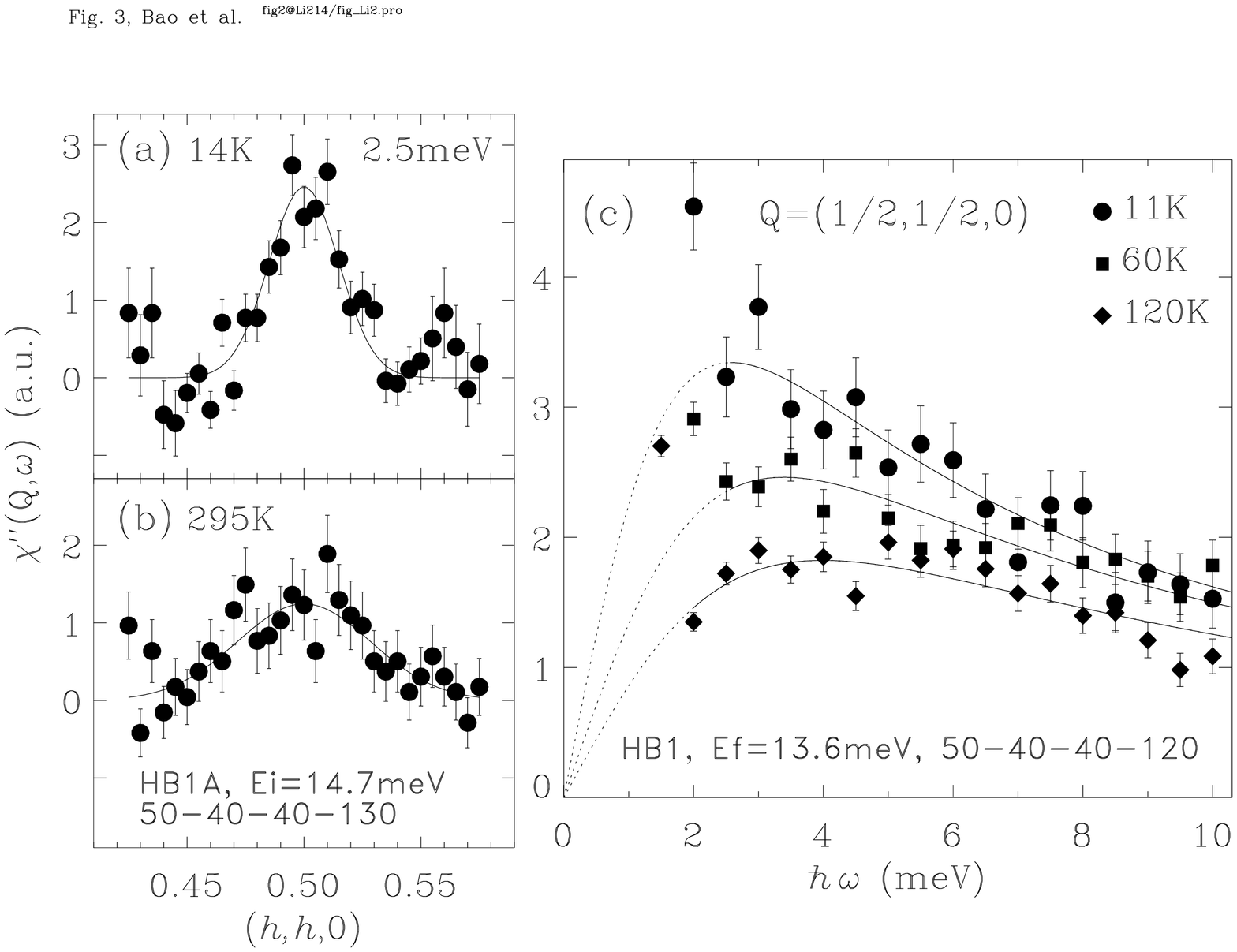,width=\columnwidth,angle=90,clip=}}
\caption{ 
Const-$E$ scans at 14~K (a) and 295~K (b). Correlation length of the
dynamic magnetic correlations is reduced by temperature, as evidenced by
a broader peak width at 295~K. The peak intensity of $\chi''({\bf Q},\omega)$
is also reduced by temperature.
(c) Dynamic magnetic susceptibility at the ($\pi,\pi$) point
as a function of energy at three temperatures. 
The temperature dependence shows the expected behavior for dynamic
magnetic correlations. 
}
\label{fig_t}
\end{figure}

Single crystals of La$_2$Cu$_{1-z}$Li$_{z}$O$_4$ were grown 
in CuO flux, using isotopically enriched $^7$Li (98.4\%) to reduce
neutron absorption. The size of single crystals grown in
this way decreases
with increasing $z$. We choose $z=0.10(2)$ for this work 
to balance the sample
size with the detectability of $\delta$, keeping in mind that $\delta$
increases with hole concentration in all other doped La$_2$CuO$_4$
materials. 
Magnetization measurements show no long-range magnetic order, 
consistent with previous studies\cite{Li214,Li214_ryu}.
The sample has orthorhombic $Cmca$ symmetry
(space group No.\ 64) at low temperatures.
In this paper, it is sufficient to use a simpler tetragonal unit
cell ($a^*=1.174 \AA^{-1}$ and $c^*=0.4814\AA^{-1}$ at 15~K) to label the 
reciprocal space; thus the $(\slantfrac{1}{2}, \slantfrac{1}{2}, 0)$ 
corresponds to the $(\pi,\pi)$ square-lattice antiferromagnetic wave vector.
A single crystal of 0.46 grams with mosaic $< 0.5^o$
was used in the {\bf Q} scans. The energy scans in Fig.~\ref{fig_t}(c)
were taken with 5 aligned crystals of total mass 1.0 grams and
mosaic of 0.9$^o$. Neutron scattering experiments were performed at
the HB1A and HB1 triple-axis spectrometers at the HFIR reactor of ORNL. 
The samples were 
mounted to the cold finger of a Displex refrigerator both in 
the $(h,k,0)$ and $(h,h,l)$ scattering planes. The spectrometer configurations 
are specified in the figures.

The intensity of neutron scattering was measured against
a neutron flux monitor placed between the sample and the exit collimator
of the monochromator. For magnetic scattering, this
intensity directly measures the
dynamic magnetic correlation function $S({\bf Q},\omega)$\cite{neut_squire},
\begin{equation}
I\propto |F({\bf Q})|^2 \cdot \overline{S}({\bf Q},\omega),
\label{eq_1}
\end{equation}
where $|F({\bf Q})|^2$ is the magnetic form factor,
and $\overline{S}({\bf Q},\omega)$ is the convolution of 
$S({\bf Q},\omega)$ with the spectrometer resolution function.
Factoring out the thermal occupation factor, 
the imaginary part of the
generalized dynamic magnetic susceptibility is given by
\begin{equation}
\chi''({\bf Q},\omega)= \left( 1-e^{-\hbar\omega/k_B T}\right)
S({\bf Q},\omega). \label{eq_fdt}
\label{eq_2}
\end{equation}
$\chi''({\bf Q},\omega)$ is a useful quantity for comparing 
the magnetic response 
at different temperatures.
Dynamic magnetic fluctuations can be approximated by a Lorentzian model,
\begin{equation}
S({\bf Q},\omega)= \frac{\hbar\omega}{1-e^{-\hbar\omega/k_B T}}\,
\frac{\chi_{\bf Q}\Gamma_{\bf Q}}
{(\hbar\omega)^2+\Gamma_{\bf Q}^2},
\end{equation}
where $\Gamma_{\bf Q}$ is the energy scale for magnetic fluctuations and
$\chi_{\bf Q}$ is the ${\bf Q}$-dependent magnetic susceptibility which
determines the intensity. The maximum of $S({\bf Q},\omega)$ is
at $\omega=0$, which is the contribution to {\em quasi}\,elastic 
scattering  from dynamic magnetic fluctuations:
\begin{equation}
S({\bf Q},0)=k_{\rm B} T\, \frac{\chi_{\bf Q}}{\Gamma_{\bf Q}}.
\end{equation}
The magnetic form factor $|F({\bf Q})|^2$ in Eq.~(\ref{eq_1})
has its maximum at ${\bf Q}=0$
while the intensity from structural excitations grows as $Q^2$. This 
fact is often used in inelastic neutron 
scattering experiments
to distinguish magnetic and structural excitations.

\begin{figure}[bt]
\centerline{
\psfig{file=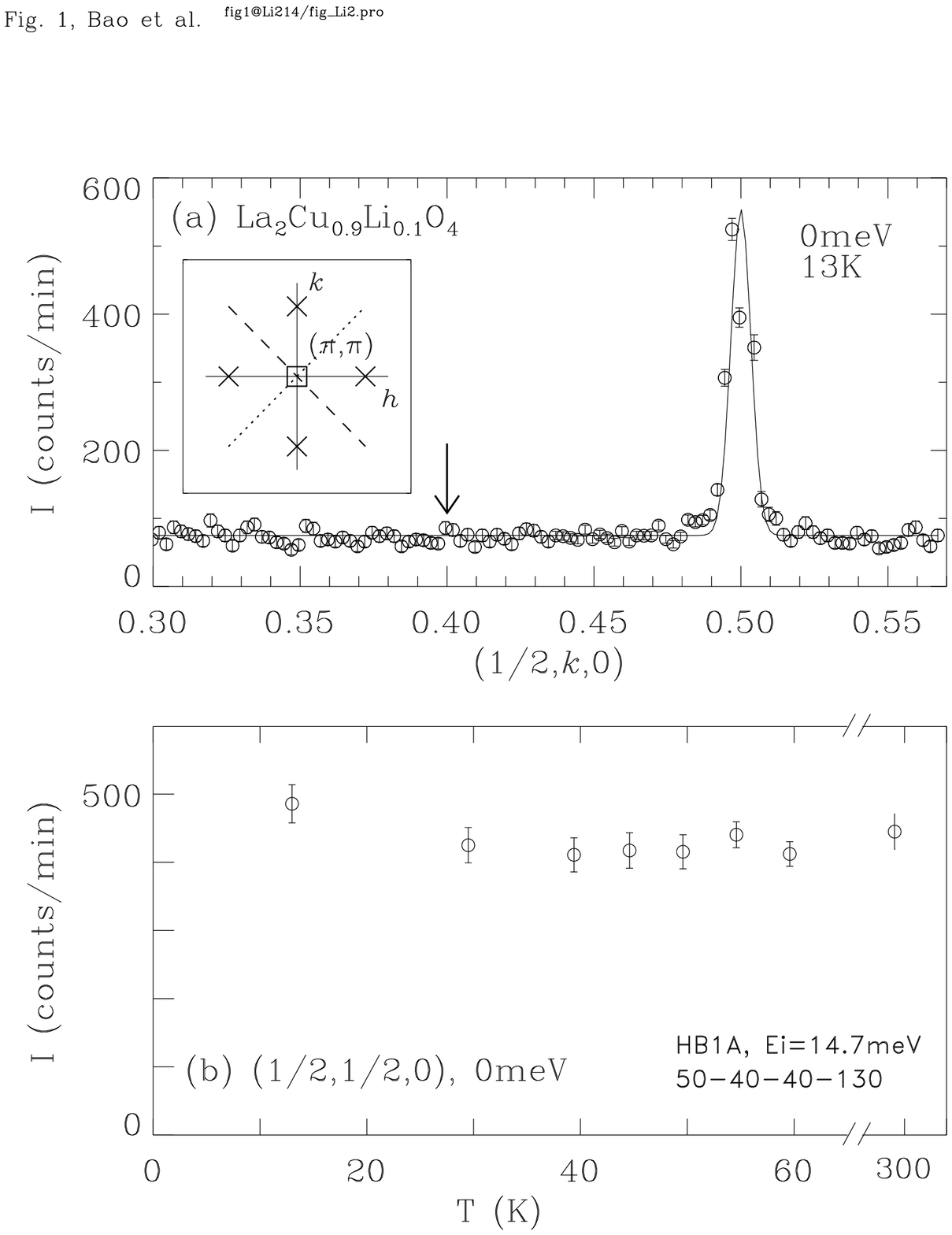,width=\columnwidth,angle=0,clip=}}
\caption{The inset in (a)
shows the 2D in-plane ($h,k$) reciprocal space. The square
marks the commensurate ($\pi,\pi$) point which characterizes dynamic 
magnetic correlations in La$_2$Cu$_{0.9}$Li$_{0.1}$O$_4$. 
The crosses mark the quartet of 
incommensurate wave vectors found for magnetic correlations
in other doped cuprates
\protect\cite{la2ch,la2hy,la2tt,L214_ym,L214_Sr,L214_Nd,L214_Zn,L214_O,ybco_dai}. 
(a) Scans across the ($\pi,\pi$) point along the $k$ direction
at $E=0$. 
The arrow indicates the incommensurate peak position found
in other cuprates
of identical hole concentration. (b) Temperature dependence of the
$(\slantfrac{1}{2}, \slantfrac{1}{2},0)$ peak intensity.}
\label{fig_q}
\end{figure}

The 2D reciprocal space for the CuO$_2$ planes near the $(\pi,\pi)$
point is shown in the inset in Fig.~\ref{fig_q}(a). The crosses 
schematically denote
incommensurate wave vectors for the low energy spin 
fluctuations previously found
in sufficiently doped La$_2$CuO$_4$ and YBa$_2$Cu$_3$O$_{6.6}$. 
A constant energy (const-$E$) scan along the $k$ direction with 
$\hbar\omega=0$ is shown by open circles in 
Fig.~\ref{fig_q}(a). The arrow indicates the incommensurate point
where a quasielastic peak in the low energy magnetic fluctuations 
is found in other doped cuprates with 10\% hole concentration. 
Apparently, La$_2$Cu$_{0.9}$Li$_{0.1}$O$_4$
is different from its peers. The only discernible peak is at the
commensurate $(\slantfrac{1}{2}, \slantfrac{1}{2},0)$, i.e.,
the $(\pi,\pi)$ point. An extended search along the ($h,h,0$) direction
has also been conducted with a similar result.

The intensity of the superlattice peak at
$(\slantfrac{1}{2}, \slantfrac{1}{2},0)$
is only $5\times10^{-4}$ of the intensity of the structural Bragg peak (110).
By adding more filters, it can be shown that the intensity at
$(\slantfrac{1}{2}, \slantfrac{1}{2},0)$ is not due to higher-order
neutron contamination. The intensity is also insensitive to neutron
energy change from 13.5 to 14.8 meV.
However, the dominant contribution to this
weak superlattice
peak is not of magnetic nature, based on its temperature
dependence [refer to Fig.~\ref{fig_q}(b)]
and {\bf Q} dependence in other Brillouin zones. 
Further investigation as to its origin is underway.

\begin{figure}[bt]
\centerline{
\psfig{file=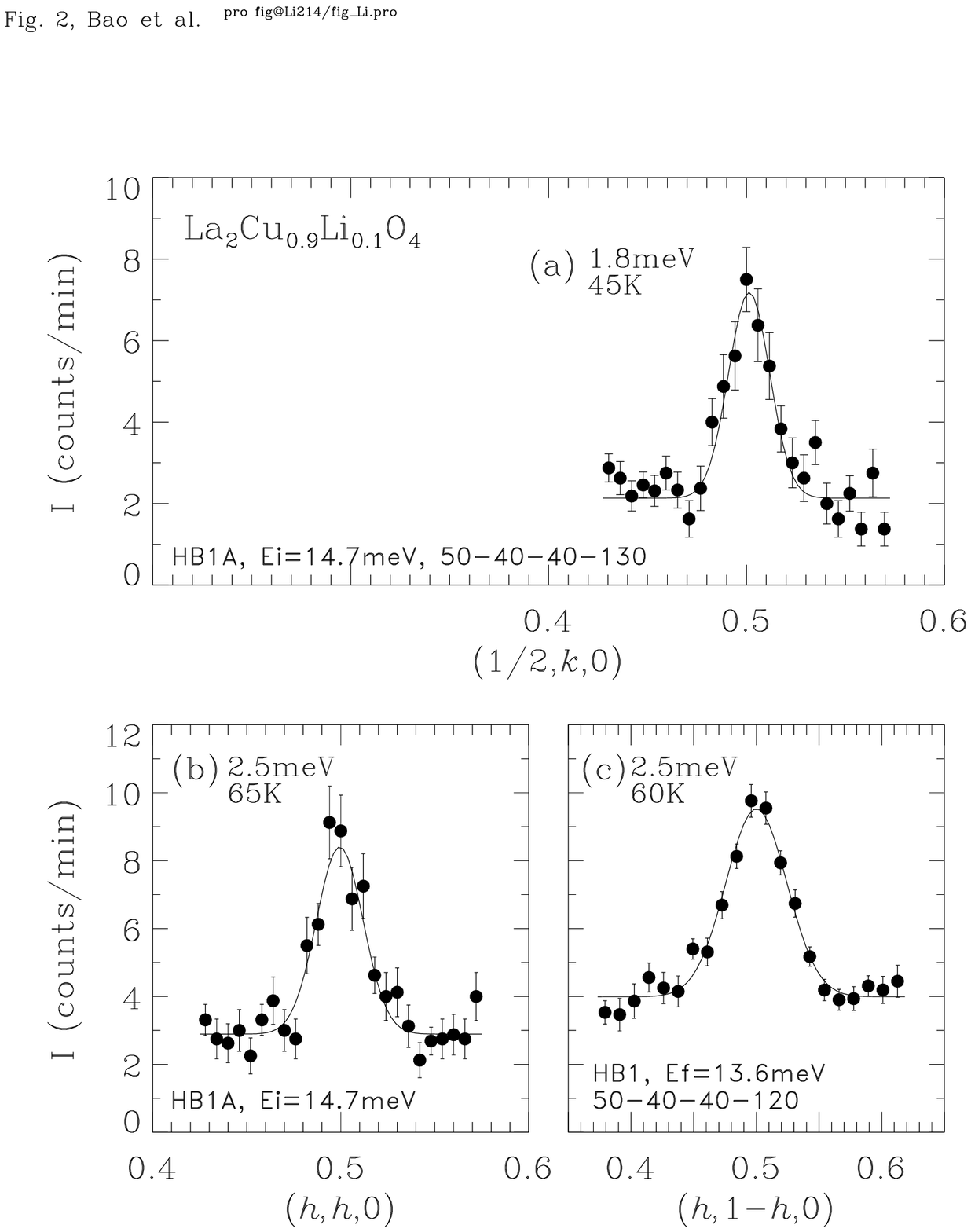,width=.9\columnwidth,angle=0,clip=}}
\caption{Const-$E$ scans across the ($\pi,\pi$) point, showing
commensurate dynamic magnetic correlations. (a) 
Scans along the $k$ direction (refer to the vertical solid line in the inset
to Fig.\ \ref{fig_q}).
(b) Scan along the ($h,h$) direction
(refer to the dotted line in the inset
to Fig.\ \ref{fig_q}).
(c) Scan along the ($h,-h$) direction (refer to the dashed line in the inset
to Fig.\ \ref{fig_q}).}
\label{fig_qe}
\end{figure}

To detect the 
dynamic magnetic correlations, 
we have repeated the $k$ scan
at a finite energy, $\hbar\omega=1.8$~meV, 
that avoids the elastic superstructure contribution. 
Results are shown
in Fig.~\ref{fig_qe}(a). 
The const-$E$ scan measures a peak at 
${\bf Q}=(\slantfrac{1}{2}, \slantfrac{1}{2},0)$ 
in the dynamic magnetic correlation function $S({\bf Q},\omega)$. 
Scans along two other symmetrically
inequivalent directions are shown in Fig.~\ref{fig_qe}(b) and (c), 
further supporting the conclusion that the magnetic correlations
are commensurate. 
Using measured phonon intensity at a similar 
energy and temperature near (110) (8 counts per minute), 
the $Q^2$ scaling factor in the phonon scattering cross-section 
and the Bragg intensity ratio between the
$(\slantfrac{1}{2}, \slantfrac{1}{2},0)$ and (110), 
the estimated acoustic phonon contribution 
near $(\slantfrac{1}{2}, \slantfrac{1}{2},0)$ is negligible.

In the left frames of Fig.~\ref{fig_t},
the dynamic magnetic correlations are compared
at 14~K and 295~K in an identical const-$E$ scan. The data have been
converted to $\chi''$, using Eq.~(\ref{eq_2}).
The shorter correlation length of the dynamic magnetic correlations
at the higher temperature is reflected in the broader peak
width at 295~K.
The energy dependence of the commensurate magnetic correlations
is shown in Fig.~\ref{fig_t}(c), with const-{\bf Q} scans at three
different temperatures. The upturning data
points above the dotted curves at low energy contain elastic
contributions due to the finite energy resolution of the spectrometer.
The energy scale, indicated by
the peak position of $\chi''$, increases with rising temperature.
In both the const-$E$ scans [Fig.~\ref{fig_t}(a) and (b)] and 
const-{\bf Q} scans [Fig.~\ref{fig_t}(c)],
the magnitude of $\chi''$ decreases with rising temperature.
All of these features are as expected for short-range dynamic magnetic order.
They reinforce the observation that the commensurate dynamic 
correlations at $(\pi,\pi)$
we found in La$_2$Cu$_{0.9}$Li$_{0.1}$O$_4$ are magnetic.

Prior to this study, a unified picture of magnetic correlations
for the La$_2$CuO$_4$ system
was emerging, namely that the incommensurability follows a 
universal function of hole concentration. This universality occurred
whether or not the material is superconducting, and whether or not
the doping was in the CuO$_2$ plane
\cite{la2ch,la2hy,la2tt,L214_ym,L214_Sr,L214_Nd,L214_Zn,L214_O}. 
The sole double-layered material in which incommensurability has so 
far been observed, YBa$_2$Cu$_3$O$_{6.6}$, also follows
the universal function for the single-layered system\cite{ybco_dai}.
Incommensurate magnetic correlations are also found in the doped
insulating nickelates\cite{Ni214_Sr} and are recently
detected in superconducting Bi$_2$Sr$_2$CaCu$_2$O$_8$\cite{bssco}.
This robust occurrence of incommensurate magnetic correlations in
charge-doped laminar materials, especially the independence on transport
properties, has been used to support the stripe picture over the nesting
Fermi surface picture, since there is no Fermi surface in the insulating
materials.
Commensurate magnetic correlations ($\delta=0$) reported here 
for La$_2$Cu$_{0.9}$Li$_{0.1}$O$_4$, which has a doped hole 
concentration of 10\%, provide a first exception to this 
empirical rule.
Another attempt to unify magnetic behavior in cuprates
is to plot the incommensurability 
versus the superconducting transition temperature. A linear relation
between them is found in La$_{2-x}$Sr$_x$CuO$_4$\cite{L214_Sr} and possibly
also in YBa$_2$Cu$_3$O$_{6+y}$\cite{sasha2}.
It is argued that this linear relation provides an important clue to the
origin of the high transition temperature superconductivity in 
cuprates\cite{sasha}. Our result on La$_2$(Cu,Li)O$_4$ may be fitted
into this scheme as a trivial limiting case: T$_C=0$ at $\delta=0$. However,
we note that measurements on (La,Sr)$_2$CuO$_4$ that is
codoped with isovalent Zn substituted for Cu
seem to violate this rule\cite{L214_Zn}.

Mobility of the doped holes may be another dimension which needs 
to be included in the picture. 
When a hole is bound to
a dopant, such as in the Li-doped case, the Skyrmion,
a long-range topological disturbance for spins surrounding the dopant, 
is favored as shown theoretically by Hass et al.\cite{hass}. 
It is expected that the long range antiferromagnetic order may be
destroyed with a dilute
Skyrmion concentration and the short range magnetic correlations
remain commensurate. This is consistent with our neutron scattering
data for Li-doped cuprate. In Zn codoped (La,Sr)$_2$CuO$_4$, holes
are introduced by Sr dopants and they are known to be more mobile.
The isovalent Zn dopants may serve as impurity pinning 
centers for the incommensurate structure formed by the holes.
It remains interesting to understand the
differences in magnetic correlations 
between La$_2$(Cu,Li)O$_4$ and hole-doped La$_2$NiO$_4$, both of which
are insulating.

In summary, we have shown by neutron scattering that dynamic 
magnetic correlations
in La$_2$Cu$_{0.9}$Li$_{0.1}$O$_4$ are commensurate with the CuO$_2$ square
lattice. This is different from all other previously 
investigated materials in the hole-doped
La$_2$CuO$_4$ and YBa$_2$Cu$_3$O$_{6+y}$ systems, which remarkably follow
a universal dependence on hole concentration. 

We wish to thank P.\ C.\ Hammel, D.\ Vaknin, K.\ Hirota, A.\ V.\ Balatsky,
S.\ A.\ Trugman, A.\ Moreo, D.\ Scalapino, A.\ L.\ Chernyshev,
J.\ M.\ Tranquada, T.\ M.\ Rice, F.\ C.\ Zhang for 
discussions and communications. W. B., R. M. and R. H. also wish to
thank J. A. Fernandez-Baca, S. Nagler, M. Yethiraj, H. Mook, B.
Chakoumakos for their hospitality at ORNL.
Work at LANL and Ames Lab.\ was conducted under the auspices
of US Department of Energy, at ORNL supported by US DOE under contract No.\
DE-AC05-960R22464 with Lockheed Martin Energy Research Corp.

\end{document}